\RequirePackage[final]{graphicx}
\documentclass[aps, superscriptaddress, prl, twocolumn, 10pt]{revtex4-2}
\usepackage[utf8]{inputenc}
\usepackage{amsmath}
\usepackage{amssymb}
\usepackage{amsfonts}
\usepackage{mathtools}
\usepackage{fixme}
\usepackage{color}
\usepackage{braket}
\usepackage{dsfont}
\usepackage{hyperref}

\usepackage{tikz}
\usetikzlibrary{calc}

\usetikzlibrary{positioning,arrows}
\usetikzlibrary{decorations}
\usetikzlibrary{decorations.pathmorphing,positioning}
\usetikzlibrary{decorations.markings}

\tikzset{square left brace/.style={ncbar=0.1cm}}
\tikzset{square right brace/.style={ncbar=-0.1cm}}

\definecolor{myred}{RGB}{214,26,70}
\definecolor{myreddark}{RGB}{76,8,38}
\definecolor{myblue}{RGB}{35,106,185}
\definecolor{mybluedark}{RGB}{19,56,99}
\definecolor{mybluebright}{RGB}{225,236,249}

\def\bk{{\bf k}}
\def\bp{{\bf p}}
\def\bq{{\bf q}}

\def\bi{{\bf i}}
\def\bj{{\bf j}}

\def\bQ{{\bf Q}}
\def\bK{{\bf K}}

\def\calG{\mathcal{G}}

\def\calP{\mathcal{P}}

\def\nn{\nonumber}

\def\AFM{{ \rm AFM }}

\def\Ham{{ \hat{H} }}
\def\dualP{{ \hat{\calP} }}

\begin{document}
\title{Dual spectroscopy of quantum simulated Fermi-Hubbard systems}
\author{K.\ Knakkergaard Nielsen}
\affiliation{Max Planck Institute of Quantum Optics, Hans-Kopfermann-Str. 1, D-85748 Garching, Germany}
\affiliation{Niels Bohr Institute, University of Copenhagen, Jagtvej 128, DK-2200 Copenhagen, Denmark}
\author{M. \ Zwierlein}
\affiliation{Department of Physics, Massachusetts Institute of Technology, Cambridge, MA 02139, USA}
\author{G.\ M.\ Bruun}
\affiliation{Department of Physics, Aarhus University, Ny Munkegade, DK-8000 Aarhus C, Denmark}
\date{\today}

\begin{abstract}
Quantum gas microscopy with atoms in optical lattices provides remarkable insights into the real space properties of many-body systems, but does not directly reveal the nature of their fundamental excitation spectrum. Here, we demonstrate that radio-frequency spectroscopy can reveal the quasi-particle nature of doped quantum many-body systems, crucial for our understanding of, e.g., high-temperature superconductors. In particular, we showcase how the existence and energy of magnetic polaron quasi-particles in doped Fermi-Hubbard systems may be probed, revealed by hallmark peaks in the spectroscopic spectrum. In combination with fundamental dualities of the Fermi-Hubbard model, we describe how these findings may be tested using several experimental platforms.
\end{abstract}

\maketitle

Quantum simulators using atoms in optical lattices provide a wealth of information regarding the spatial correlations of many-body systems at the level of single particles \cite{Bloch2012,Gross2017,Tarruell2018,Schafer2020}. This outstanding success hinges on two core capabilities: (1) single-site resolution of the atoms and their internal state is achieved using quantum gas microscopy \cite{Bakr2009,Sherson2010,Cheuk2015,Haller2015,Gross2021}, and (2) the same state is faithfully reinitialized thousands of times \cite{Tarruell2018}. One has observed antiferromagnetic correlations for the repulsive Fermi-Hubbard model in real space \cite{Greif2013,Hart2015,Boll2016,Parsons2016,Cheuk2016b,Mazurenko2017}, and charge-density wave correlations for the associated attractive model \cite{Mitra2018,Hartke2023}. Moreover, the magnetic frustration around dopants in repulsive \cite{Koepsell2019,Prichard2024,Lebrat2024} and attractive \cite{Hartke2023} Fermi-Hubbard models have been studied, and argued to stem from quasi-particle states called magnetic polarons. The non-equilibrium dynamics of a hole released from a specific lattice site has also been observed \cite{Ji2021} and successfully explained by a theory based on the formation of magnetic polarons \cite{Nielsen2022_2}. So far, there is, however, no \emph{direct} evidence for the existence of such magnetic polarons. In addition, it is debated theoretically if and when magnetic polarons form \cite{Sheng1996,Wu2008,Zhu2015,White2015,Sun2019,Zhao2022}. These questions have direct impact on the nature of pairing in such systems at higher doping, including high-temperature superconductors \cite{HighTc},  and an unambiguous evidence for the existence of magnetic polarons is paramount. 

In continuous atomic gases, radio-frequency (RF) spectroscopy has proven to be a powerful probe for the properties of Fermi \cite{Schirotzek2009,Kohstall2012,Cetina2015,Cetina2016,Scazza2017,Baroni2024} and Bose polarons \cite{Jorgensen2016,Hu2016,Yan2020,Skou2021,Skou2022,Etrych2024}, i.e.\ quasi-particles of impurities in Fermi and Bose gases. In particular, the existence of these polarons is clearly confirmed by sharp spectral peaks at their energies. In a similar fashion, angle-resolved photoemission spectroscopy (ARPES) is a key technique for mapping out single particle excitations in condensed matter systems \cite{Inglesfield:1992aa}. In optical lattices, ARPES techniques have explored excitations of the attractive Fermi-Hubbard model \cite{Brown2020a}, and recently reached a level of accuracy that makes it possible to study magnon excitations in detail \cite{Prichard2025}. In such systems, quench spectroscopy also offers an interesting pathway for spectroscopy \cite{Gritsev2007,Jurcevic2015,Menu2018,Villa2019,Villa2020,Chen2024,Bocini2025,Sun2025}. There is still, however, a dire need for exploring simple, yet powerful, spectroscopic techniques for quantum simulated many-body systems in optical lattices. 

\begin{figure}[t!]
\begin{center}
\includegraphics[width=1\columnwidth]{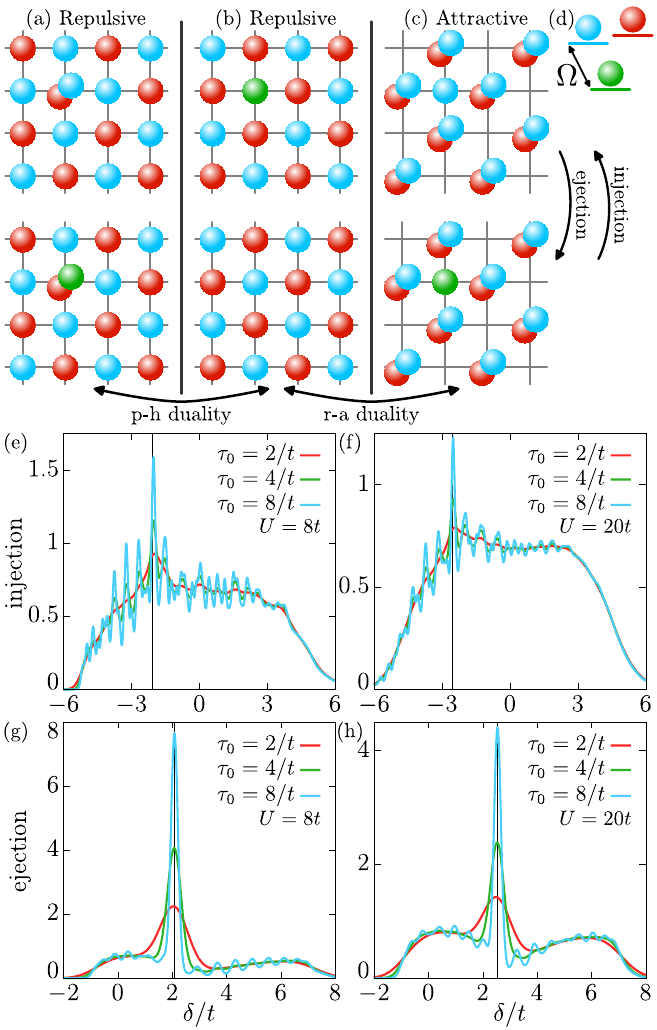}
\end{center}
\vspace{-0.5cm}
\caption{Radio-frequency spectroscopy of Fermi-Hubbard systems [red spheres: spin-$\uparrow$, blue spheres: spin-$\downarrow$, green spheres: free $\ket{f}$ state]. (a) Probing particle-doped system with repulsive interactions. (b) Probing hole-doped system with repulsive interactions. (c) Probing particle-doped system with attractive interactions. In ejection (injection) spectroscopy, the system starts out in one of the top (bottom) states and an RF transition (d) with Rabi-coupling $\Omega$ is driven. These systems are dual to each other (bidirectional arrows), up to the presence or absence of additional free particles (green). While we illustrate these processes as local flips, the probe naturally generates superpositions of dopants at any site. (e,f) Normalized transferred atoms [Eq. \eqref{eq.RFspectroscopy}] in injection spectroscopy for indicated interaction strengths $U$ and pulse lengths $\tau_0$, yielding a peak response at the quasi-particle energies (black vertical lines), but with broad spectra and poor signal-to-noise. (g,h) Normalized transferred atoms [Eq. \eqref{eq.RFspectroscopy}] in \emph{ejection} spectroscopy starting from the quasi-particle states instead singles out resonant responses at \emph{minus} the quasi-particle energy (black vertical lines) and much better signal-to-noise.}
\label{fig.spectroscopy} 
\vspace{-0.25cm}
\end{figure} 

Here, we present a careful analysis of how RF spectroscopy can be used to reveal the spectral properties of the fundamental states of the Fermi-Hubbard model. Performing microscopic calculations based on the self-consistent non-perturbative Born approximation, which is known to be quantitatively accurate even for strong interactions \cite{Martinez1991,Diamantis2021}, we show that the presence of magnetic polarons close to half-filling can be detected by the presence of a clear peak in ejection spectroscopy at a frequency determined by its energy [Fig.~\ref{fig.spectroscopy}(a-c) top to bottom], which is the hallmark feature of quasi-particles. We on the other hand find that injection spectroscopy [Figs. \ref{fig.spectroscopy}(a-c) bottom to top] has crucial limitations for detecting delocalized quasi-particles. Additionally, because of the duality between repulsive and attractive Fermi-Hubbard models on bipartite lattices \cite{Ho2009}, we show that the desired evidence for magnetic polarons can be found both for repulsive and attractive interactions alike.

\emph{Model and dualities}.-- Consider spin-$1/2$ fermions in a square lattice described by the Fermi-Hubbard model 
\begin{equation}\label{eq.Hubbard}
\!\!\Ham_{U} = \!- t\sum_{\braket{\bi,\bj},\sigma}\!\!\Big(\hat{c}^\dagger_{\bi\sigma}\hat{c}_{\bj\sigma} \!+\! {\rm H.c.}\Big) \!+\! U \!\sum_{\bi}\!\Big(\hat{n}_{\bi\uparrow} \!-\! \frac{1}{2}\Big)\!\Big(\hat{n}_{\bi\downarrow} \!-\! \frac{1}{2}\Big)\!\!
\end{equation}
where $\hat{c}^\dagger_{\bi\sigma}$ creates a spin $\sigma=\uparrow,\downarrow$ particle at lattice site $\bi$,
$\hat{n}_{\bi\sigma} = \hat{c}^\dagger_{\bi\sigma}\hat{c}_{\bi\sigma}$ is the local density operator for spin-$\sigma$, $t$ is the hopping amplitude, and $U$ is the onsite interaction strength. In the above form, the Hamiltonian implicitly incorporates a chemical potential $\mu = U / 2$, meaning that the ground state is at half filling: $\braket{\hat{n}_{\bi}} = \sum_{\sigma}\braket{\hat{n}_{\bi\sigma}} = 1$. 

The Fermi-Hubbard model  has two dualities. First, the particle-hole transformation, $\dualP_{\rm ph} \hat{c}_{\bi\sigma}\dualP_{\rm ph}^\dagger = \hat{c}_{\bi\sigma}^\dagger e^{i\bQ\cdot\bi}$, is a symmetry for any bipartite lattice, $\dualP_{\rm ph}\hat H_U\dualP_{\rm ph}^\dagger=\hat H_U$, where the phase factor $e^{i\bQ\cdot\bi}$ with $\bQ=(\pi,\pi)$ for a square lattice (unit lattice constant) ensures an alternating sign on the two sublattices. Second, the Shiba transformation \cite{Shiba1972,Moreo2007,Ho2009}, $\dualP_{\rm ra}\hat{c}_{\bi\uparrow}\dualP_{\rm ra}^\dagger = - \hat{c}_{\bi\uparrow}, \; \dualP_{\rm ra}\hat{c}_{\bi\downarrow}\dualP_{\rm ra}^\dagger = e^{i\bQ\cdot\bi}\hat{c}^\dagger_{\bi\downarrow}$, maps the repulsive model to an attractive one, $\dualP_{\rm ra}\Ham_{U}\dualP_{\rm ra}^\dagger = \Ham_{-U}$. 
Crucially, these dualities take a given hole-doped eigenstate $\ket{\Psi_n}$ for repulsive interactions, and give dual eigenstates $\dualP_{\rm ph}\ket{\Psi_n}$ ($\dualP_{\rm ra}\ket{\Psi_n}$), corresponding to a doublon-doped repulsively interacting state (spin-$\downarrow$ doped attractively interacting state).
 
\begin{figure}[t!]
\begin{center}
\includegraphics[width=1\columnwidth]{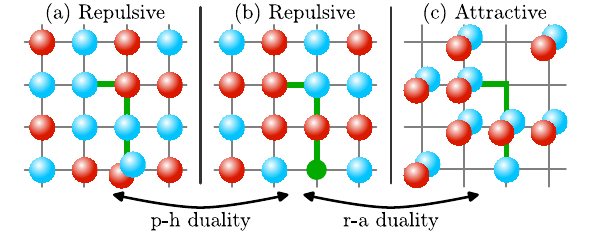}
\end{center}
\vspace{-0.5cm}
\caption{Particle-hole duality between a doublon (a) and a hole (b) moving in a repulsively interacting Fermi-Hubbard system, and the repulsive-attractive duality with an additional spin-$\downarrow$ moving in an attractively interacting system (c).}
\label{fig.duality_hole_spin_down_motion} 
\vspace{-0.25cm}
\end{figure}

To understand how these dualities may be utilized for spectroscopy, we note that the antiferromagnetically ordered N{\'e}el state, $\ket{\rm N} = \prod_{\bi\in A}\hat{c}^\dagger_{\bi\uparrow} \prod_{\bj\in B}\hat{c}^\dagger_{\bj\downarrow} \ket{0}$, approximating the ground state of the repulsive Fermi-Hubbard model at half-filling is dual to the charge-density wave 
$
\dualP_{\rm ra}\ket{\rm N} \propto \prod_{\bi\in A} \hat{c}^\dagger_{\bi\uparrow} \hat{c}^\dagger_{\bi\downarrow} \ket{0} = \ket{\rm CDW},
$
which approximates the ground state for attractive interactions \footnote{The same mapping of the N{\'e}el state in the x-y plane leads to a superfluid state (SF) \cite{Mitra2018}, showing that the CDW and SF are degenerate at half filling}. These dualities, moreover, mean that the motion of a $\downarrow$-hole (one lacking spin-$\downarrow$) just below half-filling in a repulsively interacting Fermi-Hubbard system is dual to the motion of an additional spin-down (a doublon) just above half-filling for attractive (repulsive) interactions, as illustrated in Fig. \ref{fig.duality_hole_spin_down_motion}. These dualities are convenient because different experimental scenarios may favor either particle- \cite{Koepsell2019} or hole-doping \cite{Ji2021} and attractive or repulsive interactions, as have been explored in recent proposals to measure pairing correlations \cite{Schlomer2024,Mark2024}. Here, we show how they are useful for an unambiguous spectral detection of magnetic polarons consisting of a hole or a doublon in an AFM, or equivalently a doublon in a CDW. 

\emph{RF spectroscopy.}--- Inspired by the  success of RF spectroscopy to probe quasi-particles in continuum gases \cite{Massignan2014,Massignan2025}, we now explore how it can be used to detect magnetic polarons in optical lattices. Consider a probe, 
\begin{equation} \label{eq.RF_probe}
\Ham_{\rm RF}(\tau) = \Omega(\tau)e^{-i\delta \tau}\sum_{\bk} \hat{f}^\dagger_{\bk}\hat{c}_{\bk\downarrow} + 
{\rm H.c.}, 
\end{equation}
transferring particles between the interacting $\ket{\downarrow}$ state and a third non-interacting (free) state, $\ket{f}$, at a detuning $\delta$ relative to the transition frequency between the $\ket{f}$ and $\ket{\downarrow}$ states in a vacuum. Here, $\hat{f}_{\bk}$ removes a $\ket{f}$ particle with crystal momentum $\bk$ and energy $\varepsilon^f_{\bk} = -2t(\cos k_x + \cos k_y)$. The probe, hereby, creates or removes dopants in the system as illustrated in Fig. \ref{fig.spectroscopy}. We allow the Rabi frequency $\Omega(\tau)$ to depend on time $\tau$ to model a real experiment, focusing on Gaussian pulses with pulse length $\tau_0$, such that $\Omega(\tau) = \Omega_0/(2\pi)^{1/4}\exp[-\tau^2/(4\tau_0^2)]$. Since it is experimentally challenging to simultaneously trap the three internal states $\ket{\uparrow}$, $\ket{\downarrow}$, and $\ket{f}$, we assume in the following that the initial state, before the RF probe is applied, only contains $\ket{\uparrow}$- and $\ket{\downarrow}$-particles. Linear response then gives that the number of atoms in the $\ket{f}$ state after the RF pulse is \cite{SM}
\begin{equation}\label{eq.RFspectroscopy}
N_f(\delta) = \int_{-\infty}^{\infty}\! \frac{d\omega}{2\pi} {\Omega}^2(\delta - \omega) A(\omega)
\end{equation}
with the spectral function $A(\omega)=-2{\rm Im}G(\omega)$, and 
\begin{align}\label{eq.BasicGreens}
G(\tau)&=-i\theta(\tau) \sum_\bk\bra{\Psi} \hat{c}^\dagger_{\bk\downarrow}(\tau)\hat{f}_{\bk}(\tau) \hat{f}^\dagger_{\bk}(0)\hat{c}_{\bk\downarrow}(0)\ket{\Psi} \nn \\
&= -i\theta(\tau) \sum_\bk e^{-i\varepsilon_\bk^f \tau}\bra{\Psi} \hat{c}^\dagger_{\bk\downarrow}(\tau)\hat{c}_{\bk\downarrow}(0)\ket{\Psi}
\end{align}
the Green's function, where $\hat a(\tau)=\exp(i\hat H\tau)\hat a\exp(-i\hat H\tau)$
with $\hat{f}_{\bk}(\tau)=e^{-i\varepsilon_\bk^f \tau}$ $\hat{f}_{\bk}(0)$ for the free state. Here, $\ket{\Psi}$ is the initial state containing only $\sigma =\uparrow,\downarrow$ particles. For notational simplicity, $F(\omega)$ denotes the frequency-dependent Fourier transform of the time-dependent $F(\tau)$.

\emph{Low energy description}.-- We now introduce a powerful theory for the low energy $\epsilon\ll U$ dynamics of holes close to half filling for strong repulsion $U/t\gg 1$. Integrating out doubly occupied states with high energy $\sim U$  yields the  $t$--$J$ model \footnote{The mapping also leads to small next-nearest neighbor density-dependent hopping term on the order of $J$ \cite{Auerbach_book}. While this may lead to minor quantitative changes, it should not alter the qualitative quasi-particle behavior described here.} with an AFM groundstate at half filling due to the super-exchange coupling $J = 4t^2/U$. 
Using a Holstein-Primakoff transformation generalized to include the presence of a hole (slave fermion representation), combined with linear spin wave theory, in turn gives the effective Hamiltonian
\begin{equation}\label{eq.H_eff}
\Ham_{\rm {eff}} = \sum_{\bq,\bk}g(\bq,\bk)\left[\hat{h}^\dagger_{\bq + \bk}\hat{h}_{\bq}\hat{b}^\dagger_{-\bk} + {\rm H.c.}\right]+\sum_{\bk}\omega_\bk\hat{b}^\dagger_{\bk} \hat{b}_{\bk}
\end{equation}
describing the low energy dynamics of holes in the AFM \cite{SchmittRink1988,Kane1989,Nielsen2021}. Here, $\hat{h}^\dagger_\bk$ and $\hat{b}^\dagger_{\bk}$ create a spinless holon and a spin-wave with crystal momentum $\bk$; the latter with energy $\omega_\bk = J \sqrt{1 - \gamma_\bk^2} / 2$ where 
 $\gamma_\bk = (\cos k_x + \cos k_y)/2$. 
 The spin waves are defined on top of the AFM ground state: $\hat{b}_{\bk}\ket{\AFM} = 0$. The Hamiltonian in Eq. \eqref{eq.H_eff} describes the dominant process, in which the hopping-induced scattering of a hole between crystal momenta $\bq$ and $\bq + \bk$ leads to emission/absorption of a spin-wave with momentum $-\bk$ with amplitude $g(\bq, \bk)$ \cite{SM}. While the effective Hamiltonian Eq. \eqref{eq.H_eff} is considerably simpler than Eq. \eqref{eq.Hubbard}, there is still no analytical solution available for the ground state in the presence of a single hole. On the other hand, the self-consistent Born approximation (SCBA) \cite{Kane1989,Martinez1991}, in which a subset of so-called non-crossing Feynman diagrams for the holon Green's function $\calG(\bk,\tau)= -i \theta(\tau) \bra{\AFM}\hat{h}_\bk(\tau) \hat{h}_\bk^\dagger\ket{\AFM}$ is retained, \emph{quantitatively} describes the low energy  states of the $t$-$J$ model with one hole in an AFM background \cite{Diamantis2021}. Indeed, it reproduces a ground state magnetic polaron at crystal momentum $(\pi/2,\pi/2)$ with energies in good agreement with exact diagonalization for small systems \cite{Martinez1991}. 

\emph{Injection spectroscopy.}-- Consider first the case where the initial state is the ground state AFM at half-filling, i.e. $\ket{\Psi}=\ket{\AFM}$ in Eq.\ \eqref{eq.BasicGreens}. The RF-probe then creates a hole by flipping a spin-$\downarrow$ to the free state. Since the hole can create a magnetic polaron by interacting with the AFM, it corresponds to so-called injection spectroscopy \cite{Massignan2014}. Moreover, the indicated dualities are only up to the presence or absence of the third non-interacting state. 

We can now use $\hat{c}_{\bk\downarrow} \simeq (\hat{h}^\dagger_{-\bk} - \hat{h}^\dagger_{-\bk + \bQ})/2$ to express the Green's function in Eq. \eqref{eq.BasicGreens} with the initial state $\ket{\Psi}=\ket{\AFM}$, in terms of the holon Green's function as $G(\tau)=\sum_\bk \exp(-i\varepsilon_\bk^f \tau)\calG(\bk,\tau)/2$. 
The lattice symmetry $\bQ-\bp\leftrightarrow\bp$ has here been used to simplify the expression. Figures \ref{fig.spectroscopy}(e)-\ref{fig.spectroscopy}(f) show the resulting RF injection spectrum calculated using the SCBA for various probe lengths and two different values of $U/t$. The spectra are broad, because the system starts out in an AFM state with localized particles. The RF-probe, therefore, creates holes at all crystal momenta with equal amplitude. It follows that a full continuum of states given by  $\delta=\varepsilon^{\rm pol}_{\bk}+\varepsilon^f_{\bk}$
where $\varepsilon^{\rm pol}_\bk$ is the polaron dispersion, appears in the spectral response. There is, however, a weak resonant signal remaining at the ground state polaron energy $\varepsilon^{\rm pol}_{(\pi/2,\pi/2)}$ where the density of states is large. Nevertheless, these results demonstrate that injection spectroscopy suffers from detrimental broadening obscuring a clear spectral observation of magnetic polarons. A workaround is to post-select  the  momentum of the free state \cite{Bohrdt2018}, which, however, complicates an already involved experimental sequence. 

\emph{Ejection spectroscopy.}--
To circumvent this broadening  and avoid challenging post-selection, we propose instead to use ejection spectroscopy. Here, one starts out in the state of interest, i.e., a system with a single (or low density of) dopant(s) forming magnetic polarons. It is not favorable, however, to probe a repulsively interacting hole-doped system, because the initial hole is formed by a non-interacting $\ket{f}$ particle, requiring the simultaneous trapping and cooling of all three internal states as illustrated in Fig. \ref{fig.spectroscopy}(b). We instead consider the case where the initial state $\ket{\Psi_{\bp}}$ contains one doublon, i.e., one extra $\downarrow$-particle, forming a magnetic polaron with momentum $\bp$ on top of the AFM (or CDW) background for repulsive and attractive interactions respectively, see Figs. \ref{fig.spectroscopy}(a) and \ref{fig.spectroscopy}(c). Using the particle-hole or repulsive-attractive dualities discussed above, the Green's function in Eq. \eqref{eq.BasicGreens} can be expressed as $G(\bp,\tau)=-i\theta(\tau)\sum_\bk\exp(-i\varepsilon_\bk^f\tau)\bra{\tilde\Psi_{\bp}}\hat c_{\bk\downarrow}(\tau)\hat c^\dagger_{\bk\downarrow}(0)\ket{\tilde\Psi_{\bp}}$, where the state $\ket{\tilde\Psi_{\bp}}=\dualP_{\rm ph}\ket{\Psi_{\bp}}$ (or $\dualP_{\rm ra}\ket{\Psi_{\bp}}$) contains one $\downarrow$-hole on top of an AFM. This explicitly shows that ejection spectroscopy starting from a doublon in an AFM/CDW for repulsive/attractive interaction is equivalent to ejection spectroscopy starting with a $\downarrow$-hole in an AFM for repulsive interaction as illustrated in Fig. \ref{fig.spectroscopy}. 

In addition to its experimental importance, this duality is useful theoretically, as we can use the  SCBA to calculate the wave function $\ket{\tilde\Psi_{\bp}}$ of the magnetic polaron formed by a hole in the AFM. Indeed, one can expand $\ket{\tilde\Psi_\bp}$ in terms containing an increasing number of spin waves created by the hole as 
\begin{align}
\ket{\Psi^{\rm pol}_\bp} = \sqrt{Z_{\bp}}(\hat{h}^\dagger_\bp + \sum_{\bk} \phi_{\bp,\bk}\hat{b}^\dagger_{-\bk}\hat{h}^\dagger_{\bp+\bk} 
+\dots)\ket{\AFM}
\label{eq.quasiparticle_state}
\end{align}
where $Z_\bp$ is the quasi-particle residue of the magnetic polaron, $\phi_{\bp,\bk} = g(\bp,\bk) \calG(\bp \!+\! \bk,\varepsilon^{\rm pol}_\bp \!-\! \omega_\bk)$ are the expansion parameters set by the SCBA \cite{Reiter1994,Ramsak1998}, and $\ldots$ denote terms with more spin waves. In this picture, the Green's function is expressed as $G(\bp,\tau)=-i\theta(\tau)\sum_\bk\exp(-i\varepsilon_\bk^f\tau)\bra{\Psi^{\rm pol}_{\bp}}\hat{h}_{\bk}^\dagger(\tau)\hat{h}_{\bk}(0)\ket{\Psi^{\rm pol}_{\bp}}$, describing the dominant response by which the hole is refilled, and ignoring doublon production separated by the large energy scale $U$ \footnote{Note that $\ket{\Psi^{\rm pol}_{\bf p}}$ has vanishing mean magnetization, $\braket{\hat{S}^{(z)}} = 0$, because holes in this state are uniformly created on all lattice sites, whereas $\ket{\tilde{\Psi}_{\bf p}}$ has magnetization $+1/2$. Since the system is symmetric in spin-up and -down, they give equivalent spectral responses.}. To illustrate the main physics, we first approximate the polaron wave function with the dominant first term $\ket{\tilde\Psi_\bp} \sim \sqrt{Z_{\bp}}\hat{h}^\dagger_\bp\ket{\AFM}$. This gives $G(\bp,\tau) \sim Z_\bp\exp[i(\varepsilon^{\rm pol}_\bp - \varepsilon_\bp^f)\tau]$. Using this in Eq. \eqref{eq.RFspectroscopy} then gives a sharp peak in the RF-response for 
\begin{equation}\label{eq.RFpeakEjection}
\delta=\varepsilon^{f}_{\bp}-\varepsilon^{\rm pol}_\bp.
\end{equation}
The interpretation of Eq.\ \eqref{eq.RFpeakEjection} is straightforward: the RF probe has to provide energy to flip the magnetic polaron with energy $\varepsilon^{\rm pol}_\bp$ into a free particle with energy $\varepsilon^{f}_{\bp}$. Hence, observing this sharp peak will confirm the presence of a magnetic polaron and it is a major advantage of using ejection spectroscopy as compared to injection spectroscopy. Using the SCBA, we can derive exact expressions for the expansion coefficients in Eq. \eqref{eq.quasiparticle_state} to all orders in the number of spin waves \cite{Reiter1994,Ramsak1998}, resulting in
\begin{align} \label{eq.ejection_greens_expansion}
\!\!G(\bp,\delta) =& Z_\bp\Bigg[\frac{1}{\delta \!+\! \varepsilon^{\rm pol}_\bp \!-\! \varepsilon^f_\bp \!+\! i\eta} \nn \\
&+ \sum_{\bk} \frac{[g(\bp,\bk)\calG(\bp \!+\! \bk,\varepsilon^{\rm pol}_\bp \!-\! \omega_{-\bk})]^2}{\delta \!+\! \varepsilon^{\rm pol}_\bp \!-\! \omega_{-\bk} \!-\! \varepsilon^f_{\bp+\bk} \!+\! i\eta} \!+\! \dots \Bigg],\!\!
\end{align}
with the positive infinitesimal $\eta = 0^+$. Besides the resonant peak given by Eq. \eqref{eq.RFpeakEjection}, Eq. \eqref{eq.ejection_greens_expansion} shows that a continuum, 
$
\delta = \varepsilon^f_{\bp + \sum_{i=1}^n\bk_i} + \sum_{i=1}^n \omega_{-\bk_i} - \varepsilon^{\rm pol}_\bp
$
is present in the RF response corresponding to the probe removing the magnetic polaron and creating $n$ spin waves alongside the free particle. To circumvent having to calculate Eq. \eqref{eq.ejection_greens_expansion} order by order, we make use of the fact that the SCBA endows the wave function with a self-similar structure, which has previously been used to derive self-consistency equations for correlation functions \cite{Nielsen2021,Nielsen2022_2}. Indeed, by introducing an auxiliary function $\Gamma(\bp,\delta;\omega)$ that is the self-consistent solution to
\begin{align}
\!\!\!\!&\Gamma(\bp,\delta;\omega) = \frac{1}{\delta \!+\! \omega \!-\! \varepsilon^f_\bp \!+\! i\eta} + \nn\\
\!\!\!\!&\sum_{\bk} [g(\bp,\bk)\calG(\bp \!+\! \bk,\omega \!-\! \omega_{-\bk})]^2 \Gamma(\bp \!+\! \bk,\delta;\omega \!-\! \omega_{-\bk}),\!\!
\end{align}
we compute the full Green's function as: $G(\bp,\delta) = Z_\bp \Gamma(\bp,\delta; \omega = \varepsilon^{\rm pol}_\bp)$ \cite{SM}, allowing us to include an \emph{infinite} number of spin-wave terms in Eq. \eqref{eq.quasiparticle_state}. 

The resulting ejection spectra [Figs. \ref{fig.spectroscopy}(g) and \ref{fig.spectroscopy}(h)] are plotted at the same probe lengths and values of $U/t$ as the injection spectra in Figs. \ref{fig.spectroscopy}(e) and \ref{fig.spectroscopy}(f), 
assuming the initial state to be the ground state polaron with momentum $\bp=(\pi/2,\pi/2)$. As predicted, the resonant signal given by Eq. \eqref{eq.RFpeakEjection} is much clearer than in the injection spectra. Unlike what is often the case in ARPES and RF spectroscopy, the many-body continuum forms both below and above this resonance. This is a result of three properties: (1) the free state is formed at the variable momentum $\bp + \sum_{i=1}^n\bk_i$, (2) the free state has a large energetic bandwidth of $8t$, and (3) the polaron ground state at $\bp = (\pi/2,\pi/2)$ is in the middle of that bandwidth, $\varepsilon^f_{\bp} = 0$. Figures \ref{fig.spectroscopy}(g) and \ref{fig.spectroscopy}(h) moreover show that the spectral weight of the continuum increases with $U/t$ at the expense of the quasi-particle peak. This happens, because the system becomes more strongly interacting with increasing $U/t$, corresponding to decreasing $J/t$, so that the residue, $Z_\bp$, of the polaron decreases.

\begin{figure}[t!]
\begin{center}
\includegraphics[width=1\columnwidth]{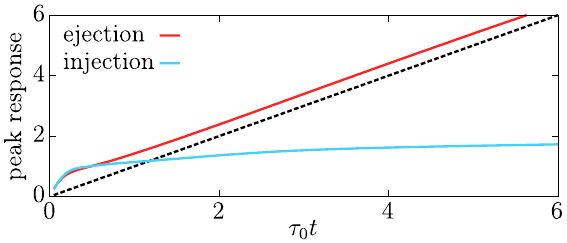}
\end{center}
\vspace{-0.5cm}
\caption{Peak response for injection [units: $Nt/(\Omega^2_0\tau_0)$] (blue) and ejection spectroscopy [units: $t/(\Omega^2_0\tau_0)$] (red) vs pulse length, $\tau_0$, for $U = 8t$. While the former quickly saturates, the latter increases linearly (dashed black line).}
\label{fig.peak_response} 
\vspace{-0.25cm}
\end{figure}

Finally, we make the principle advantage of ejection spectroscopy even more clear by calculating how the peak response, i.e. the height of the quasi-particle peak in the spectra, depends on the pulse length $\tau_0$ [Fig. \ref{fig.peak_response}]. This demonstrates that the ejection spectroscopy peak scales linearly $\sim Z_\bp\tau_0$ with pulse length, whereas the inferior injection peak response quickly saturates. 

\emph{Experimental considerations.}-- In addition to a finite pulse length, $\tau_0$, any RF-experiment has frequency fluctuations. Writing the total frequency as $\delta+\Delta$, and assuming that the additional noise $\Delta$ follows a normal distribution around $\Delta = 0$ effectively leads to a shorter probing pulse, $\tilde{\tau}_0 = (\tau_0^{-2} + 4\sigma_\Delta^2)^{-1/2}$ \cite{SM}. Here, $\sigma_\Delta$ is the standard deviation in the probe frequency. In the absence of fluctuations, $\sigma_\Delta = 0$, Fig. \ref{fig.spectroscopy} establishes that we need pulse lengths $>1/t$ for a clear signal. Including fluctuations, $\sigma_\Delta \neq 0$, in turn means that we need $\tilde{\tau}_0 > 1/t$, or put differently: for a certain
pulse length $\tau_0$, the required precision on the probe frequency is $\sigma_\Delta \lesssim (t^2 - \tau_0^{-2})^{-1/2}/2$. More
concretely, the attractive Fermi-Hubbard setup in Ref.\ \cite{Hartke2023} has a hopping amplitude of $t \simeq 2\pi \times 350 \, {\rm Hz}$. Realistically probing the quasi-particle then requires $\tau_0 \gtrsim \frac{2}{t} \simeq 0.9 \,{\rm ms}$ and $\sigma_\Delta \lesssim 2\pi \times 150 \,{\rm Hz}$. These explicit estimates show that our RF ejection spectroscopic scheme of the magnetic polaron can realistically be implemented in current experimental setups. The desired accuracy on the order of the hopping amplitude has indeed recently been achieved for probing dopant-dressed magnons in ferromagnetic backgrounds \cite{Prichard2025}. More generally, our proposed RF  scheme requires three internal states of the same atomic species and to find a Feshbach resonance for which \emph{only} the  interaction between two states, $\ket{\uparrow}$ and $\ket{\downarrow}$, is significant. The ability to \emph{freely} choose between attractively and repulsively interacting systems may, therefore, be instrumental for actual implementations. An alternate setup could use \emph{two} layers, in which a second empty layer is used for detection, and where the tunneling between the layers plays the role of the RF probe in Eq. \eqref{eq.RF_probe} \cite{Bohrdt2018,Priv_Comm_Thomas_Pohl}. 

\emph{Conclusions.}-- Using a non-perturbative theory for the Fermi-Hubbard model close to half-filling, we systematically analysed how its low energy spectrum can be explored using RF ejection spectroscopy with atoms in optical lattices. In particular, we demonstrated that a smoking gun detection of magnetic polarons is possible via the presence of a clear quasi-particle peak in the RF ejection spectrum. Moreover, we showed that the particle-hole and repulsive-attractive dualities of the Fermi-Hubbard model makes our scheme realistic for current quantum simulators using both repulsive \cite{Koepsell2019,Ji2021} and attractive \cite{Hartke2020} interactions. Finally, we demonstrated how pulse lengths longer than the inverse hopping, $1/t$, and probing frequency accuracies on the order of $100 \, {\rm Hz}$ are sufficient to clearly resolve the quasi-particle peaks, bringing the experimental realization within reach. Our probing scheme generally investigates transitions between  states with $N_d$ and $N_d - 1$ dopants, and, therefore, naturally extends to investigating many-body states in general such as for example bound magnetic polaron pairs.

\begin{acknowledgments}
K.K.N. acknowledges support from the Carlsberg Foundation through a Carlsberg Reintegration Fellowship (Grant no. CF24-1214). G.M.B. acknowledges the Danish National Research Foundation through the Center of Excellence CCQ (Grant no. DNRF156). M.Z. acknowledges support by the NSF CUA and PHY-2012110 and AFOSR (FA9550- 23-1-0402).
\end{acknowledgments}

\emph{Data availability.} The data that support the findings of this article are openly available \cite{data_availability}.

\bibliographystyle{apsrev4-2}
\bibliography{ref_dual_spectroscopy}

\end{document}


\title{Supplemental Material \\ Dual spectroscopy of quantum simulated Fermi-Hubbard systems }
\maketitle
\beginsupplement
\tableofcontents

\section{Linear response}
In this section, we give more details on the derivation of the linear response for spectroscopy. We start from the Hamiltonian for the radio-frequency transfer in the rotating wave approximation
\begin{equation}
\Ham_{\rm RF}(\tau) = \Omega(\tau)e^{-i\delta \tau} \sum_{\bk} \hat{f}^\dagger_{\bk}\hat{c}_{\bk\downarrow} + {\rm H.c.},
\end{equation}
where $\delta = \omega_{\rm pr} - (\varepsilon^{(f)} - \varepsilon^{(\uparrow)})$ is the detuning of the probe frequency $\omega_{\rm pr}$ to the bare transition frequency $\varepsilon^{(f)} - \varepsilon^{(\uparrow)}$. For simplicity, we assume that the pulse envelope $\Omega(\tau)$ is real and even in $\tau$. We also assume that the non-interacting state $\ket{f}$ is trapped, and may move with the same hopping amplitude as the spin-$\ket{\uparrow}$ and -$\ket{\downarrow}$ states
\begin{equation}
\Ham_f = -t\sum_{\braket{\bi,\bj}} \left[\hat{f}^\dagger_{\bi}\hat{f}_{\bj} + {\rm H.c.}\right] = \sum_{\bk} \varepsilon^{f}_\bk \hat{f}^\dagger_{\bk}\hat{f}_{\bk},
\end{equation}
with $\varepsilon^{f}_\bk = -2t[\cos k_x + \cos k_y]$. Let us now see, how the linear response signal arises. For both injection and ejection spectroscopy, we assume that the initial state only contains $\ket{\uparrow}$- and $\ket{\downarrow}$-particles. We are then interested in computing the total production, $N_f$, in the non-interacting third state, $\ket{f}$, as a function of detuning $\delta$, as the probe is turned on. This is directly observable in quantum simulation experiments by utilizing spin-specific single-site resolution. Alternatively, at the end of the experimental run, one can push out all atoms remaining in the $\ket{\uparrow},\ket{\downarrow}$-states and then count the remaining atoms in the $\ket{f}$-state. The general linear response formula for an operator $\hat{A}$, 
\begin{align}
\braket{\hat{A}} = \braket{\hat{A}}_0 - i\int_{-\infty}^\tau {\rm d}s \braket{\big[(\hat{A})_I, (\hat{V}(s))_I \big]}_0,
\label{eq.linear_response}
\end{align}
subject to a time-dependent perturbation $\hat{V}(s)$ can now be used to calculate what we are after. Here, $\braket{\cdot}_0$ means an average with respect to the non-perturbed [$\hat{V} = 0$] state. In particular, we set $\hat{A} = \partial_\tau \hat{N}_f$ [$\hat{N}_f = \sum_\bk \hat{f}^\dagger_\bk \hat{f}_\bk$], and $\hat{V}(s) = \Ham_{\rm RF}(s)$. Here, a sub-index $I$ means that we are calculating observables in the interaction picture with $\hat{V} = \Ham_{\rm RF}$ the interaction Hamiltonian. Hence,
\begin{align}
(\partial_\tau \hat{N}_f)_I &= -i \left[\hat{N}_f, \Ham_{\rm RF}(\tau)\right] = -i \Omega(\tau) \sum_{\bk,\bq}\left[\hat{f}^\dagger_\bk(\tau) \hat{f}_\bk(\tau), e^{-i\delta \tau}\hat{f}^\dagger_{\bq}(\tau)\hat{c}_{\bq\downarrow}(\tau) + e^{+i\delta \tau}\hat{c}^\dagger_{\bq\downarrow}(\tau)\hat{f}_{\bq}(\tau)\right] \nn \\
&= -i\Omega(\tau)\sum_\bk \left( e^{-i\delta\tau} \hat{f}^\dagger_\bk(\tau) \hat{c}_{\bk\downarrow}(\tau) - e^{+i\delta\tau} \hat{c}^\dagger_{\bk\downarrow}(\tau) \hat{f}_\bk(\tau)\right),
\end{align}
with $\hat{f}_\bk(\tau) = e^{+i\Ham\tau}\hat{f}_\bk e^{-i\Ham\tau} = e^{-i\varepsilon_\bk^{f}\tau} \hat{f}_\bk$, $\hat{c}_{\bk\downarrow}(\tau) = e^{+i\Ham\tau}\hat{f}_{\bk\downarrow} e^{-i\Ham\tau}$, and $\Ham = \Ham_{t,U} + \Ham_f$. Inserting this into Eq. \eqref{eq.linear_response} together with a bit of rearranging yields
\begin{align}
&\braket{\partial_\tau \hat{N}_f} = \Omega(\tau) \int_{-\infty}^\tau ds \,\Omega(s) \sum_{\bk,\bq} \Big(e^{i(\delta - \varepsilon^{f}_\bk)\tau} e^{-i(\delta - \varepsilon^{f}_\bq)s} \braket{\big[ \hat{c}^\dagger_{\bk\downarrow}(\tau) \hat{f}_{\bk}, \hat{f}^\dagger_{\bq} \hat{c}_{\bq\downarrow}(s) \big]}_0 + {\rm c.c.}\Big). 
\label{eq.linear_response_calc_1_ejection}
\end{align}
Since there are no non-interacting $\ket{f}$-atoms in the initial state, we get that 
\begin{align}
\braket{\big[ \hat{c}^\dagger_{\bk\downarrow}(\tau) \hat{f}_{\bk}, \hat{f}^\dagger_{\bq} \hat{c}_{\bq\downarrow}(s) \big]}_0 = \delta_{\bq,\bk} \braket{ \hat{c}^\dagger_{\bk\downarrow}(\tau) \hat{c}_{\bq\downarrow}(s)}_0 = \delta_{\bq,\bk} \braket{ \hat{c}^\dagger_{\bk\downarrow}(\tau - s) \hat{c}_{\bq\downarrow}(0)}_0,
\end{align}
Finally, we define $G(\tau) = -i\theta(\tau)\sum_\bk e^{-i\varepsilon^f_\bk \tau} \braket{\hat{c}^\dagger_{\bk\downarrow}(\tau) \hat{c}_{\bk\downarrow}(0)}_0$. Using this above then shows that 
\begin{align}
N_f(\delta) &= \int_{-\infty}^{+\infty}d\tau \braket{\partial_\tau \hat{N}_f} = \int_{-\infty}^{+\infty}d\tau\,\Omega(\tau) \int_{-\infty}^{+\infty} ds \,\Omega(s) \Big(e^{i\delta(\tau-s)} i G(\tau-s) + {\rm c.c.}\Big) \nn \\
&= \int_{-\infty}^{+\infty}\frac{d\omega}{2\pi}\,\Omega(\delta- \omega)\Omega(\omega-\delta) A(\omega) = \int_{-\infty}^{+\infty}\frac{d\omega}{2\pi}\,\Omega^2(\delta - \omega)A(\omega).
\label{eq.linear_response_calc_2_ejection}
\end{align}
The third equality follows from Fourier transformation, $f(\tau) = \int_{-\infty}^{+\infty}d\omega \,e^{-i\omega\tau}f(\omega)/(2\pi)$, and using $A(\omega) = -2{\rm Im}G(\omega)$. The fourth and last equality follows from the fact that since $\Omega(\tau)$ is even in $\tau$, i.e., even in the time domain, $\Omega(\omega)$ is even in $\omega$, i.e., even in the frequency domain. This proves Eq. (4) of the main text. 

\section{Self-consistent Born approximation and injection Green's function}
The starting point is the coupling of the hole to the linear spin waves, $\Ham^{\rm eff}$, in Eq. (6) of the main text. From here, the holon Green's function, $\calG(\bp,\tau) = -i\theta(\tau)\bra{\AFM}\hat{h}_\bp(\tau)\hat{h}^\dagger_\bp(0)\ket{\AFM}$ is approximated by taking only the rainbow diagrams for the self-energy into account \cite{Kane1989,Martinez1991,Nielsen2021}
\begin{equation}
\Sigma(\bp,\omega) = \sum_\bk g^2(\bp,\bk) \calG(\bp,\omega - \omega_\bk),
\label{eq.hole_self_energy}
\end{equation}
where $[\calG(\bp,\omega)]^{-1} = \omega - \Sigma(\bp,\omega) + i0^+$. Also, $g(\bp, \bk) = 4t[u_\bk\gamma_{\bp + \bk} - v_\bk\gamma_\bp]/\sqrt{N}$, where $N$ is the number of lattice sites, and $u_\bk = [(J/(2\omega_\bk) + 1)/2]^{1/2}, v_\bk = {\rm sgn}(\gamma_\bk)[(J/(2\omega_\bk) - 1)/2]^{1/2}$ are the coherence factors of the linear spinwave theory, while $\gamma_\bk = [\cos k_x + \cos k_y]/2$ is the structure factor. Upon numerically solving Eq. \eqref{eq.hole_self_energy} for a $N = 20\times20$ square lattice, we obtain the Green's function for injection as $G(\tau) = \sum_\bk \calG(\bk,\tau)/2$, which is inserted into Eq. (4) of the main text to calculate $N_f(\delta)$ for injection spectroscopy.

\section{Ejection Green's function}
We start from the ejection Green's function for the hole
\begin{align}
\!\!G(\bp, \tau) = -i\theta(\tau)\sum_\bk e^{-i\epsilon^{\rightarrow}_\bk\tau} \bra{\Psi^{\rm pol}_\bp} \hat{h}^\dagger_\bk(\tau) \hat{h}_\bk(0) \ket{\Psi^{\rm pol}_\bp},
\end{align}
where $\ket{\Psi^{\rm pol}_\bp}$ is the magnetic polaron eigenstate in the SCBA [see Eq. (7) of the main text]. Moreover, letting $\tilde{G}(\bp,\delta) = G(\bp,\delta)/Z_\bp$, we find an expansion for $\tilde{G}$ in terms of the number of generated spin waves in the polaronic state. Namely, 
\begin{align}
\tilde{G}(\bp,\delta) &= \tilde{G}_{0}(\bp,\delta) \!+\! \sum_{\bk_1} [g(\bp,\bk_1)\calG(\bK_1, \varepsilon^{\rm pol}_\bp \!-\! \omega_{\bk_1})]^2 \times\tilde{G}_{0}(\bK_1,\delta \!-\! \omega_{\bk_1}) + \dots
\label{eq.G_ej_expansion}
\end{align}
This is equivalent to Eq. (9) of the main text. Here, $[\tilde{G}_{0}(\bp,\delta)]^{-1} = \delta + \varepsilon^{\rm pol}_\bp - \varepsilon^{\rightarrow}_\bp + i0^+$, and $\varepsilon^{\rm pol}_\bp$ is the polaron energy at crystal momentum $\bp$. Now, promoting $\varepsilon^{\rm pol}_\bp$ to a general frequency $\omega$ \cite{Nielsen2021}, and letting $\tilde{G}_{0}(\bp,\delta),\tilde{G}_{h}(\bp,\delta) \to \Gamma_{0}(\bp,\delta; \omega), \Gamma(\bp,\delta;\omega)$ then shows that 
\begin{align}
\Gamma(\bp,\delta;\omega) &= \Gamma_{0}(\bp,\delta;\omega) \!+\! \sum_{\bk_1} [g(\bp,\bk_1)\calG(\bK_1, \omega \!-\! \omega_{\bk_1})]^2 \times\Gamma_{0}(\bK_1,\delta;\omega-\omega_{\bk_1}) + \dots \nn \\
&= \Gamma_{0}(\bp,\delta;\omega) \!+\! \sum_{\bk_1} [g(\bp,\bk_1)\calG(\bK_1, \omega \!-\! \omega_{\bk_1})]^2 \times\Gamma(\bK_1,\delta;\omega-\omega_{\bk_1}),
\label{eq.G_ej_self_consistent}
\end{align}
yielding a self-consistency equation for $\Gamma(\bp,\delta;\omega)$. Solving this, the hole ejection Green's function is then evaluated as $G(\bp,\delta) = Z_\bp\Gamma(\bp, \delta; \omega = \varepsilon^{\rm pol}_\bp)$. 

\section{Fluctuations in probing frequency}
In this Section, we analyze the effect of fluctuations in the radio-frequency probe frequency, and hence in the detuning $\delta$. Consequently, we let $\Delta$ denote the additional detuning  (so $\delta + \Delta$ is the total detuning) and assume that it varies according to a normal distribution around $\Delta = 0$ between different experimental runs. Averaging over these experimental runs for the ejection spectroscopic result in Eq. \eqref{eq.linear_response_calc_2_ejection}, means that $G(\tau) \to h(\tau) G(\tau)$, with
\begin{align}
h(\tau) &= \int_{-\infty}^{+\infty} d\Delta \, e^{+i\Delta\tau} \frac{1}{\sqrt{2\pi \sigma_\Delta^2}}e^{-\Delta^2/[2\sigma_\Delta^2]} = \int_{-\infty}^{+\infty} d\phi \, e^{+i\phi} \frac{1}{\sqrt{2\pi (\sigma_\Delta\tau)^2}}e^{-\phi^2/[2(\sigma_\Delta\tau)^2]} = e^{-(\sigma_\Delta\tau)^2/2}. 
\end{align}
A similar analysis was performed in Ref. \cite{Skou2021}. In combination with the finite-pulse result in Eq. (8) leads to the double convolution 
\begin{align}
N_f(\delta) = \int_{-\infty}^{+\infty} \frac{d\omega}{2\pi}\int_{-\infty}^{+\infty} \frac{d\nu}{2\pi} \,\Omega^2(\delta - \nu) h(\nu - \omega) A(\omega),
\end{align}
where $h(\omega) = \sqrt{2\pi}/\sigma_\Delta \times \exp[-\omega^2/(2\sigma_\Delta^2)]$ is the Fourier transform of $h(\tau)$. Assuming a Gaussian pulse shape with pulse length $\tau_0$ means that $\Omega^2(\nu) = \sqrt{8\pi} (\Omega_0\tau_0)^2 \exp[-2(\tau_0\nu)^2]$. We can then carry through the $\nu$ integration above, yielding
\begin{align}
N_f(\delta) = \int_{-\infty}^{+\infty} \frac{d\omega}{2\pi} F(\delta - \omega) A(\omega),
\end{align}
with the effective pulse function, 
\begin{align}
F(\omega) = \sqrt{8\pi}\Omega^2_0 \tau_0 \times\tilde{\tau}_0 \exp[-2(\tilde{\tau}_0\omega)^2],
\end{align}
and the effective probing time
\begin{align}
\tilde{\tau}_0 = [\tau_0^{-2} + 4\sigma_\Delta^2]^{-1/2}.
\end{align}
It is, hereby, this effective probing time that needs to be $\gtrsim 1/t$. In other words, for a specific probing time $\tau_0 \geq 1/t$, the RF probing frequency needs to be stable within a deviation of $\sigma_\Delta \lesssim \sqrt{t^2 - \tau_0^{-2}}/2$. If this is not the case, the quasiparticle signal is washed out.

\bibliographystyle{apsrev4-2}
\bibliography{ref_dual_spectroscopy}